\documentclass[conference]{IEEEtran}
\IEEEoverridecommandlockouts

\usepackage{cite}
\usepackage{amsmath,amssymb,amsfonts}
\usepackage{algorithmic}
\usepackage{graphicx}
\usepackage{textcomp}
\usepackage{xcolor}
\def\BibTeX{{\rm B\kern-.05em{\sc i\kern-.025em b}\kern-.08em
    T\kern-.1667em\lower.7ex\hbox{E}\kern-.125emX}}

\usepackage{amsmath,times}
\usepackage{color}
\usepackage[dvipsnames]{xcolor}
\usepackage{amssymb}
\usepackage{tabularray}

\usepackage[mode=buildnew]{standalone}
\usepackage{multirow}
\usepackage{booktabs}
\usepackage{etoolbox, siunitx}
\usepackage{tabularx}
\usepackage{tikz}
\usepackage{pgfplots}
    \pgfplotsset{compat=1.8}

\usetikzlibrary{dsp}
\usetikzlibrary{chains}
\usetikzlibrary{shapes.geometric}
\usetikzlibrary{calc}
\usetikzlibrary{arrows.meta}
\usepackage[input-decimal-markers={,}, 
     output-decimal-marker = {,},]{siunitx}

\usepackage[normalem]{ulem}

\usepackage{cleveref}
    
\begin{document}

\title{KNOWLEDGE DISTILLATION \\ FOR EFFICIENT ACOUSTIC ECHO CONTROL}

\author{\IEEEauthorblockN{\textit{Ernst Seidel$^{\ast}$, Pejman Mowlaee$^{\circ}$, Tim Fingscheidt$^{\ast}$}}\\
\IEEEauthorblockA{$^{\ast}$Institute for Communications Technology,
	Technische Universität Braunschweig\\
	Schleinitzstraße 22,
	38106 Braunschweig, Germany\\ $^{\circ}$GN Audio A/S,
	Lautrupbjerg 7,
	2750 Ballerup, Denmark \\
\tt\small \{e.seidel, t.fingscheidt\}@tu-bs.de;
pmowlaee@gn.com}
}

\newcommand{\EV}[1]{E\{{#1}\}}  

\maketitle

\begin{abstract}

In recent years, many efforts have been made to supersede classical acoustic echo control (AEC) algorithms with more powerful machine-learned approaches. While surpassing the performance of well-established adaptive filters is very much possible, a remaining challenge is computational complexity. Popular architectures, such as convolutional recurrent networks (CRNs), are by multiple orders of magnitude computationally more expensive than classical signal processing solutions. Scaling down such models is usually straight-forward, but it comes at the cost of a notably reduced performance.
We show---to the author's knowledge for the first time in AEC---how these performance drops can be successfully alleviated to a large degree by employing an effective knowledge distillation (KD) process, enabling more potent efficient AEC. Our proposed CGGN16 student AEC models show significantly less near-end speech distortion at only 2\% of its teacher's computational complexity, surpass the overall performance of a six times more complex model trained on ground-truth labels, and outperform other AEC-focused architectures from recent literature.

\end{abstract}


\IEEEpeerreviewmaketitle

\section{Introduction}
\label{sec:intro}

\IEEEPARstart{W}{hen} communicating via a hands-free system (speakerphone), the echo of the loudspeaker signal is being picked up by the microphone, degrading severely the perceived quality of the conversation. In recent times, it has been shown that significant performance improvements of acoustic echo control (AEC) over classical adaptive filter-based approaches \cite{Feuer1985,Benesty2001,Enzner2006} can be achieved by either augmentation with a neural network \cite{Zhang2022d,Zhang2023,Yang2023,Haubner2024,Seidel2024a} or complete replacement with such. A popular architecture among the latter are convolutional recurrent networks (CRNs) \cite{Zhang2019h,Seidel2021,Braun2022,Indenbom2023}, which are often employed in the form of acoustic echo suppression (AES) models estimating a mask to be applied to the microphone signal, contrary to the classical approach of estimating and subtracting the echo.

A major drawback of employing deep neural networks (DNNs) for AEC is the much higher computational complexity and memory requirements compared to classical signal processing (SP) solutions. The large boost in computational power and memory capacity of DNNs with their millions of parameters and billions of floating point operations per second (FLOPS) is still too demanding even for modern edge devices such as conference microphones. Accordingly, we have seen a noticeable trend towards more efficient DNN architectures \cite{Pfeiffenberger2020,Chen2023,Indenbom2023,Seidel2023}.

However, decreasing the size of a DNN beyond a certain point often comes with a compromise in performance and accuracy. For an AEC system, this potentially means more residual echo and degradation of the near-end speaker's signal. An approach often seen in literature to mitigate performance loss from model downscaling is the concept of knowledge distillation (KD) \cite{Hinton2015,Yang2023a}, also known as teacher-student learning. In KD, a usually much larger and more powerful teacher model is used to train the target student model, augmenting or replacing traditional loss functions based on ground truth. While KD improves performance in fields such as noise suppression \cite{Watanabe2017,Kim2021a,Park2024,Nathoo2024}, there have been no works
on the application of KD to the AEC task.

In this work, we employ a significantly downscaled version of a convolutional grouped GRU network ({\tt CGGN16})~\cite{Seidel2023} and show how the performance drop can be significantly mitigated by employing knowledge distillation. We conduct an ablation study on various loss formulations and propose a strategy to achieve performance on par with significantly larger model variants that are solely trained on ground truth.

The remainder of this paper is structured as follows: Section 2 introduces the processing framework, baseline architecture, and the KD approach. The datasets, training, and evaluation are discussed in Section 3. Section 4 provides conclusions.

\section{System Overview and Proposed Method}
\label{sec:method}

\begin{figure}[t]
	\centering
    	\usetikzlibrary{dsp}
	\usetikzlibrary{chains}
	\usetikzlibrary{shapes.geometric}
	\usetikzlibrary{calc}

\begin{tikzpicture}[scale = 1.5]

\tikzstyle{encdec}=[trapezium, trapezium angle=67.5, draw, inner ysep=5pt, outer sep=0pt,text width=1, minimum height=10, line width=.8pt, fill=white]

\pgfarrowsdeclare{dsparrow}{dsparrow}
{
	\arrowsize=0.40pt
	\advance\arrowsize by .5\pgflinewidth
	\pgfarrowsleftextend{-4\arrowsize}
	\pgfarrowsrightextend{4\arrowsize}
}
{
	\arrowsize=0.40pt
	\advance\arrowsize by .5\pgflinewidth
	\pgfsetdash{}{0pt} 
	\pgfsetmiterjoin	 
	\pgfsetbuttcap		 
	\pgfpathmoveto{\pgfpoint{-4\arrowsize}{2.5\arrowsize}}
	\pgfpathlineto{\pgfpoint{4\arrowsize}{0pt}}
	\pgfpathlineto{\pgfpoint{-4\arrowsize}{-2.5\arrowsize}}
	\pgfpathclose
	\pgfusepathqfill
}

\pgfmathsetmacro{\toprow}{2.2}
\pgfmathsetmacro{\bottomrow}{0.9}
\pgfmathsetmacro{\middlerow}{(1.60}
\pgfmathsetmacro{\boxheight}{2}
\pgfmathsetmacro{\boxwidth}{5.5}

    \draw   (7.45, \middlerow-0.075) node    (backdrop)     [dspfilter, minimum width=13.5 em, minimum height=8.75 em, fill=black!10, draw=white] {}
            (8.5, \toprow + 0.2) node (Sys) {Near-end}
            (2.5, \toprow + 0.2) node (Sys) {Far-end}
            (2.5, \toprow)    coordinate  (InX)   {}
            (4.0, \toprow)    node  (StopX) [dspnodefull] {}

            (6.25, \toprow)  node (LS) [encdec, rotate=90] {}

            (6.8, \middlerow)     node (MidD)   [dspfilter, minimum width=3 em, minimum height=1.5 em, fill=white] {$h(n)$}
            (8.15, \middlerow+0.14)     node (rir) {room impulse}
            (8.2, \middlerow-0.14)     node (rir) {response (RIR)}
            (7.25, \bottomrow)     coordinate (S)      {}
            (6.8, \bottomrow-0.25)   coordinate (N)      {}
            
            (2.5, \bottomrow) coordinate  (OutE)  {}
            (4.0, \bottomrow) node        (GS) [dspfilter, minimum width=\boxwidth em, minimum height=\boxheight em, fill=red!25]{AEC}
            (6.25, \bottomrow) node  (MicIn) [circle, inner sep= 4, fill=white]  {}
            (6.25, \bottomrow) node  (Mic) [dspmultiplier, minimum width=0.5, fill=white]  {}
            (6.41, \bottomrow) coordinate  (InM)  {}
            (6.41, \bottomrow+0.2) coordinate  (MT)  {}
            (6.41, \bottomrow-0.2) coordinate  (MB)  {};

    \draw (4, \toprow+0.2) node (Xt) {$x(n)$};
    \draw (2.5, \bottomrow+0.3) node (Et) {$e(n)$};
    \draw (5.35, \bottomrow+0.3) node (Yt) {$y(n)$};
    \draw ([xshift=+3mm, yshift=0mm] S.east)                node (St) {$s(n)$};
    \draw ([xshift=+3mm, yshift=0mm] N.east)           node (Nt) {$n(n)$};
    \draw ([xshift=+3mm, yshift=-2.0mm] MidD.south)           node (Dt) {$d(n)$};
    \draw ([xshift=+2mm, yshift=+4.5mm] MidD.north)           node (X_t) {$x'(n)$};
    \draw ([yshift=-.5mm] InM) coordinate (InMb) {};

    \begin{scope}[start chain]
	
        \chainin (InX);
        \chainin (LS) 	[join=by dspconn];

        \chainin (Mic);
        \chainin (GS) 	[join=by dspconn];
        \chainin (OutE) 	[join=by dspconn];

        \chainin (StopX);
        \chainin (GS) 	[join=by dspconn];

        \chainin (S);
        \chainin (InM) 	[join=by dspconn];
        \chainin (N);
        \chainin (InMb) 	[join=by dspconn];

        \chainin (MT);
        \chainin (MB) [join=by dspline];

    \end{scope}

    \draw[dspconn] (LS.south) to[out=-2,in=95] (MidD);
    \draw[dspconn] (MidD) to[out=-90,in=30] ([yshift=.5mm] InM);
	
\end{tikzpicture}
	\caption{Overview of the acoustic echo control framework.}
	\label{fig:generalsystem}
\end{figure}
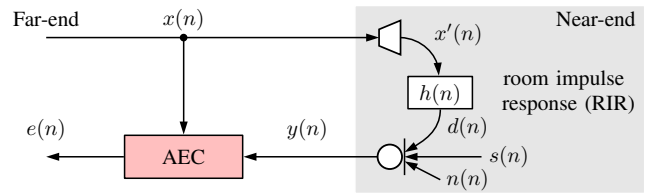

\subsection{Processing Framework}
\label{ssec:subhead}

Figure \ref{fig:generalsystem} shows the processing framework in which we employ various AEC algorithms. At our simulated near-end environment, the far-end reference signal $x(n)$ is played by a loudspeaker and propagates through the room, forming the echo signal $d(n)=x'(n)*h(n)$, with $*$ being a convolution and $h(n)$ being the room impulse response. Note that we consider loudspeaker nonlinearities by $ x'(n) = f_\mathrm{NL}(x(n))$. The echo is picked up by our microphone alongside the near-end speech $s(n)$ and background noise $n(n)$, forming our microphone signal $y(n) = s(n)+n(n)+d(n)$. The input signals to our AEC are $x(n)$ and $y(n)$.
The sampling rate of all signals is $16$\,kHz. 
As all presented models operate in the frequency domain, the input signals $y(n)$ and $x(n)$ are split into frames of $N_T = 1024$ samples with a frame shift of \mbox{$R=128$} samples. These frames are then subject to an oversampled filterbank after \cite{Harteneck1999} \mbox{(with oversampling factor of 2)} to form the respective frequency-domain signals $X_\ell(k)$ and $Y_\ell(k)$ with frame index $\ell$, frequency bin index \mbox{$k \in \mathcal{K} = \{0,1,...,K/2\}$}, and DFT length $K=512$. 
For the DNN-based AEC models, these signals are divided into their real and imaginary parts, which then constitute two input channels of the networks. All models estimate (either) a mask (or an echo estimate) which is multiplied to (subtracted from) $Y_\ell(k)$ to obtain the enhanced signal $E_\ell(k)$. Transformation back into the time domain including overlap-add yields the final output signal $e(n)$, resulting in an overall algorithmic latency of $40$\,ms.

\subsection{Baseline Architecture}
\label{ssec:architecture}

We employ the {\tt CGGN16}~\cite{Seidel2023} as our baseline AEC model. This model typically employs \mbox{$U=3$} convolutional encoder-decoder blocks, each of which compresses the size of its input feature maps by a factor 2, while expanding the number of feature maps up to $UF$ feature maps at the bottleneck. The parameter $F$ denotes the base number of feature maps from which all layer configurations are derived (cf.\ \cite{Seidel2023}).

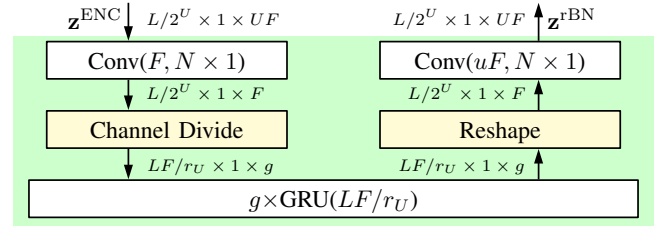
\begin{figure}[t]
    \hspace{-4mm}
    	\usetikzlibrary{dsp}
	\usetikzlibrary{chains}
	\usetikzlibrary{shapes.geometric}
	\usetikzlibrary{calc}
	
\newcommand{\QuadConv}{EDBlock}

\begin{tikzpicture}[scale = 0.75]

\pgfarrowsdeclare{dsparrow}{dsparrow}
{
	\arrowsize=0.50pt
	\advance\arrowsize by .5\pgflinewidth
	\pgfarrowsleftextend{-4\arrowsize}
	\pgfarrowsrightextend{4\arrowsize}
}
{
	\arrowsize=0.50pt
	\advance\arrowsize by .5\pgflinewidth
	\pgfsetdash{}{0pt} 
	\pgfsetmiterjoin	 
	\pgfsetbuttcap		 
	\pgfpathmoveto{\pgfpoint{-4\arrowsize}{2.5\arrowsize}}
	\pgfpathlineto{\pgfpoint{4\arrowsize}{0pt}}
	\pgfpathlineto{\pgfpoint{-4\arrowsize}{-2.5\arrowsize}}
	\pgfpathclose
	\pgfusepathqfill
}

\tikzstyle{arrow} = [thick,->,>=stealth]
\pgfmathsetmacro{\rootx}{0}
\pgfmathsetmacro{\rooty}{0}

\pgfmathsetmacro{\topconvs}{-1}
\pgfmathsetmacro{\botconvs}{-5.05}
\pgfmathsetmacro{\convsep}{1.35}

\pgfmathsetmacro{\leftrow}{-0.75}
\pgfmathsetmacro{\rightrow}{7.25}
\pgfmathsetmacro{\middlerow}{(\rightrow+\leftrow)/2}
\pgfmathsetmacro{\boxheight}{1.55}
\pgfmathsetmacro{\boxwidth}{10}
\pgfmathsetmacro{\quadboxwidth}{25.15}

    \fill [green!20] (-3.0,-4.575) rectangle (9.5,-8.275);
             \draw   
              (\middlerow,{\topconvs-3*\convsep}) coordinate    (BN) 

              (\leftrow+.75,{\botconvs-0*\convsep}) node    (bottIn)     [dspfilter, minimum width=\boxwidth em, minimum height=\boxheight em, fill=white] {Conv($F, N\times 1$)}
	       (\rightrow-0.75,{\botconvs-0*\convsep}) node    (bottOut)     [dspfilter, minimum width=\boxwidth em, minimum height=\boxheight em, fill=white] {Conv($uF, N\times 1$)}
		   (\leftrow,{\botconvs-(0.55*\convsep)})   coordinate    (bottIn_)    {}
		   (\rightrow,{\botconvs-(0.55*\convsep)})   coordinate    (bottOut_)    {}
              (\leftrow+.75,{\botconvs-1*\convsep}) node    (clstm1)     [dspfilter, minimum width=\boxwidth em, minimum height=\boxheight em, fill=yellow!20] {Channel Divide}
	       (\rightrow-0.75,{\botconvs-1*\convsep}) node    (clstm2)     [dspfilter, minimum width=\boxwidth em, minimum height=\boxheight em, fill=yellow!20] {Reshape}
              ({\middlerow},{\botconvs-2*\convsep}) node    (rr)     [dspfilter, minimum width=2.535*\boxwidth em, minimum height=\boxheight em, fill=white] {$g\times$GRU($LF/r_U$)};
              
    \draw ([xshift=-7.5mm] bottIn.south) coordinate (conv3.out);
	\draw ([xshift=-7.5 mm] bottIn.north) coordinate (conv3.in);
	\draw ([xshift=-7.5 mm] clstm1.north) coordinate (chdiv.in);

    \draw ([xshift=-7.5mm, yshift=15mm] bottIn.south) coordinate (conv2.out);

    \draw ([xshift=7.5mm] bottOut.south) coordinate (convd3.in);
	\draw ([xshift=7.5 mm] bottOut.north) coordinate (convd3.out);

	\draw ([xshift=7.5 mm] clstm2.north) coordinate (resh.out);
    \draw ([xshift=7.5mm, yshift=15mm] bottOut.south) coordinate (convd2.in);

    \draw ([xshift=-7.5 mm] clstm1.south) coordinate (chdiv.out);
    \draw ([xshift=+7.5 mm] clstm2.south) coordinate (resh.in);

    \draw ([xshift=-40 mm] rr.north) coordinate (rr.in);
    \draw ([xshift=+40 mm] rr.north) coordinate (rr.out);


    \draw ([xshift=9mm, yshift=-1mm] bottIn.north) coordinate (bottIn_data_in);
    \draw ([xshift=7.5mm, yshift=-2.0mm] clstm1.north) coordinate (clstm1_data_in);
    \draw ([xshift=10mm, yshift=0mm] clstm1.south) coordinate (clstm1_data_out);
    \draw ([xshift=-6.5mm, yshift=-2.0mm] clstm2.north) coordinate (clstm2_data_out);
    \draw ([xshift=-10mm, yshift=0mm] clstm2.south) coordinate (clstm2_data_in);
	\draw ([xshift=-8mm, yshift=-1mm] bottOut.north) coordinate (bottOut_data_out);	

    \draw  (bottIn_data_in)   node    (out_data_node)       [label=above:\scriptsize $L/2^U \times 1 \times UF$] {};
    \draw  (clstm1_data_in)   node    (out_data_node)       [label=above:\scriptsize $L/2^U\times 1 \times F$] {};
    \draw  ([xshift=-1.5mm, yshift=+1.5mm]clstm1_data_out)   node    (clstm1_out)       [label=below:\scriptsize $LF/r_U \times 1 \times g$] {};
    \draw  ([xshift=2.5mm, yshift=+1.5mm]clstm2_data_in)   node    (clstm1_out)       [label=below:\scriptsize $LF/r_U \times 1 \times g$] {};
    \draw  (clstm2_data_out)   node    (out_data_node)       [label=above:\scriptsize $L/2^U \times 1 \times F$] {};
	\draw  (bottOut_data_out)   node    (out_data_node)       [label=above:\scriptsize $L/2^U \times 1 \times UF$] {};

    \draw ([xshift=-14mm, yshift=4.25mm] bottIn.north) node (bottIn_label) {${\bf z}^\mathrm{ENC}$};
    \draw ([xshift=+14mm, yshift=4.25mm] bottOut.north) node (bottIn_label) {${\bf z}^\mathrm{rBN}$};


    \begin{scope}[start chain]
    
    	\chainin (conv2.out); 

    	\chainin (conv3.in) 		[join=by dspconn];
    	\chainin (conv3.out);

        \chainin (chdiv.in) 		[join=by dspconn];
    	\chainin (chdiv.out);

        \chainin (rr.in) 		[join=by dspconn];
    	\chainin (rr.out);

        \chainin (resh.in) 		[join=by dspconn];
    	\chainin (resh.out);

        \chainin (convd3.in) 	[join=by dspconn];
        \chainin (convd3.out);
        \chainin (convd2.in) 	[join=by dspconn];

    \end{scope}
	
\end{tikzpicture} 
    \caption{The recurrent bottleneck of the employed \texttt{CGGN16} model adopted from \cite{Seidel2023}, using a grouped GRU strategy with $g$ parallel layers (\mbox{$r_U = 2^U g$}).}
    \label{fig:CGGN}
\end{figure}

At the recurrent bottleneck after encoder/decoder block $U=3$ (shown in Fig.\ \ref{fig:CGGN}), a convolutional layer (kernel size $N=3$) reduces the feature map count of the encoder output ${\bf z}^\mathrm{ENC}$ back to $F$. These input feature maps are then divided into $g$ groups, each being processed by a separate gated recurrent unit (GRU). As such, each GRU has only $LF/r_U$ inputs and hidden units, balancing performance and overall impact on parameter count. After GRU output combination and decompression into $uF$ feature maps via another convolutional layer, the bottleneck output ${\bf z}^\mathrm{rBN}$ is passed to the decoder.

All variants of the {\tt CGGN} in this work share this architecture. We adjust the parameter count and complexity by changing $F$ and $g$. The starting point and teacher for later knowledge distillation (cf.\ Sec.\ \ref{ssec:KD}) will be a model with $F=64$, as preliminary experiments showed no performance gain from increasing the base kernel count further. The grouping parameter is generally chosen as $g=F/4$, which yields a large parameter count reduction without significant performance trade-offs.

\subsection{Proposed Knowledge Distillation Losses}
\label{ssec:KD}

While our {\tt CGGN16} was already an attempt to decrease the model size and computational complexity, the explored architecture adjustments were constrained by the goal of maintaining performance. In contrast, this work aims to drastically decrease model size and complexity and regain lost performance via knowledge distillation (KD). 
We evaluate several KD approaches, distinguished by their loss function described in the following. Our goal is to train a powerful teacher network (superscript (\,)$^\mathrm{T}$) and to use its enhanced output signal $e^\mathrm{T}_b(n)$ as a target for guiding the smaller student network (superscript (\,)$^\mathrm{S}$) towards a better performance.

As baseline ground-truth (GT) loss training without KD, we use the time-domain logMSE loss defined as
\begin{equation}
    \begin{split}
        {J}^{\mathrm{GTt}}_b & = {J}^{\mathrm{logMSE}}\big({\bf e}^\mathrm{S}_b,{\bf s}_b+{\bf n}_b\big) \\
        & =  10\!\cdot\!\log \Big( \frac{1}{N}\sum_{n \in \mathcal{N}} \big|e^\mathrm{S}_b(n)-s_b(n)-n_b(n)\big|^2  \Big),
    \end{split}
    \label{eq:GTt}
\end{equation}
over the entire time sequence \mbox{$\big(n \in \mathcal{N}=\{0, ..., N\!-\!1\}\big)$}, before averaging over all batch entries $b \in \mathcal{B}=\{1, ..., B\}$, with $e^\mathrm{S}_b(n)$ being the enhanced output signal of the student network. Note that with \eqref{eq:GTt} and in the context of all our experiments, we only aim at echo suppression, not noise reduction.

For the first attempt in employing knowledge distillation, the teacher output replaces the GT signal in the loss.
Both GT loss~\eqref{eq:GTt} and KD logMSE loss can be combined in a joint loss
\begin{equation}
    {J}^{\mathrm{GTKDt}}_b = \alpha {J}^{\mathrm{GTt}}_b + (1-\alpha) {J}^{\mathrm{logMSE}}\big({\bf e}_b^\mathrm{S},{\bf e}_b^\mathrm{T}\big)
    \label{eq:GTKDt}
\end{equation}
with weighting parameter $\alpha=0.5$. In analogy to this time-domain approach, the KD loss in the frequency domain reads
\begin{equation}
    \begin{split}
        {J}^\mathrm{KDf}_b 
        & = 10\!\cdot\!\log \Big(\frac{1}{LK}\sum_{\ell \in \mathcal{L}}\sum_{k \in \mathcal{K}} \big|E^\mathrm{S}_{b,\ell}(k)-E^\mathrm{T}_{b,\ell}(k)\big|^2  \Big)
    \end{split}
    \label{eq:KDf}
\end{equation}
for frame index $\ell\in \mathcal{L}=\{1, ..., L\}$ and frequency bin index \mbox{$k\in \mathcal{K}=\{0, ..., K\!\!-\!\!1\}$}. Our resulting joint loss analog to \eqref{eq:GTKDt} still uses the time-domain GT loss~\eqref{eq:GTt} and $\alpha=0.5$:
\begin{equation}
    {J}^{\mathrm{GTKDf}}_b = \alpha {J}^{\mathrm{GTt}}_b + (1-\alpha) {J}^\mathrm{KDf}_b.
    \label{eq:GTKDf}
\end{equation}
 
\section{Experimental Evaluation and Discussion}
\label{sec:setup}

\subsection{Datasets and Training Details}
\label{ssec:training}

We adopt the training and test setup detailed in our previous work \cite{Seidel2024}, including parameterization of the datasets $\mathcal{D}_{\mathrm{train}}$, $\mathcal{D}_{\mathrm{con}}$, $\mathcal{D}_{\mathrm{dev}}$, and $\mathcal{D}_{\mathrm{test}}$. The dev set $\mathcal{D}_{\mathrm{dev}}$ is designed to be close to (but disjoint from) the training data $\mathcal{D}_{\mathrm{train}}$, while the {\it test set $\mathcal{D}_{\mathrm{test}}$ uses completely different resources and parameterization to test our models' generalization capabilities in unseen conditions}. Speakers for this test set are taken from the TIMIT corpus~\cite{TIMIT}, noise types from the ETSI noise database ~\cite{ETSI2008}, and nonlinearities are modeled by the arctan nonlinearity function~\cite{Jung2013, Zoellzer2003}. The RIRs are sampled from the Aachen Impulse Response Database \cite{Jeub2010}, provides real-world recordings from different acoustic environments. Signal components are mixed at a signal-to-echo ratio (SER) and signal-to-noise ratio (SNR) chosen from $\{-9, -6, ..., 9\}$\,dB and $\{5, 8, ..., 20\}$\,dB, respectively.
\textit{All data is publicly available online}, whereby only TIMIT requires a small fee.

Our scripts for data generation and evaluation as well as implementations of the frequency-domain Kalman filter ({\tt FDKF}) and {\tt CGGN16} models are provided in our software toolbox \cite{Seidel2024b}. The {\tt DLAC-Kalman} is adopted from the toolbox provided by \cite{Haubner2024}. {\tt CRUSE-AEC} is implemented based on the author's description in \cite{Braun2022}. All models are causal and trained from scratch on a {\tt GTX 1080 Ti} {\rm GPU} using {\tt PyTorch2}~\cite{Paszke2019}. Training runs are deterministic (fixed seed), using the Adam optimizer~\cite{Kingma2015} with a batch size of $16$ and a backpropagation-through-time unrolling sequence length of $200$ frames. The learning rate (LR) starts at $10^{-4}$ and is halved after 10 epochs without loss improvement on the control split $\mathcal{D}_{\mathrm{con}}$. If the LR drops below $10^{-5}$ or if the loss on $\mathcal{D}_{\mathrm{con}}$ does not improve for 20 consecutive epochs, training is stopped.

\subsection{Evaluation Metrics}
\label{ssec:metrics}

Results are mainly discussed on the challenging double-talk (DT) condition, although the single-talk far-end (STFE) and single-talk near-end (STNE) conditions have been evaluated during experiments as well. Each condition is evaluated on its own subset of metrics. For some metrics, we employ their black-box variants (marked by the subscript $(\,)_\mathrm{BB}$), utilizing individual components of the enhanced signal \mbox{$e(n)=\tilde{s}(n)+\tilde{d}(n)+\tilde{n}(n)$} according to ITU-T Recommendations P.1100~\cite{ITU-P1100} and P.1110~\cite{ITU-P1110}, with more details in \cite{Fingscheidt2007, Fingscheidt2008}. This allows for a more in-depth analysis with focus on the specific effects of residual echo and NE speech preservation. Echo control is measured using the black-box echo return loss enhancement (ERLE$_\mathrm{BB}$) metric after \cite{Fingscheidt2007, Vary2006}, defined as
\begin{equation}
    \mathrm{ERLE_\mathrm{BB}}(n)=10\cdot\mathrm{log}_{10}\left(\frac{\EV{d^2(n)}}{\EV{\big(\tilde{d}^2(n)\big)}}\right) \quad \text{in dB},
    \label{eq:ERLE}
\end{equation}
with black-box echo component $\tilde{d}(n)$, the expectation operator $\EV{\cdot}$ approximated by a first-order IIR smoothing filter with impulse response $g(n)=\alpha^n, n\in\{0,1,2,...\}$, and coefficient \mbox{$\alpha=0.99$}. The final ERLE$_\mathrm{BB}$ is computed as mean over the entire evaluated sequence. Preservation of NE speech is gauged by the
PESQ metric~\cite{ITU_P862.2_Corr1} in both its standard \mbox{PESQ($e(n),s(n)$)} and black-box implementation \mbox{PESQ$_\mathrm{BB} =$ PESQ($\tilde{s}(n),s(n)$)}, as well as
the log-spectral distance (LSD) \cite{Katsir2011} averaged over all frames $\ell$ of a file \textit{that contain near-end speech.}
The Levenshtein phone similarity (LPS) is employed for measurement of the NE speaker's phonetic fidelity in signal $e(n)$, with ${\rm LPS} = 1 - {\rm LPD}$ and the Levenshtein phone distance (LPD) from \cite{Pirklbauer2023}. 
We also report speech intelligibility by ESTOI~\cite{Taal2016}.
Additionally, the \mbox{AECMOS} Echo metric~\cite{purin2021aecmos} in DT and single-talk (labeled DT\,E and ST\,E) and AECMOS Other in DT (DT\,O) are employed to instrumentally estimate MOS scores regarding echo control effectiveness and overall NE speech quality, respectively. We further conducted a crowd-sourced P.808/P.831 subjective listening test after \cite{ITU_P808, ITU_P831, Cutler2021a}, delivering MOS scores labeled DT\,O* (NE speech quality) and DT\,E* (echo annoyance) on a noiseless variant of $\mathcal{D}_{\mathrm{test}}$ to verify the acoustical significance of our findings.

Both $\mathcal{D}_{\mathrm{dev}}$ and $\mathcal{D}_{\mathrm{test}}$ allow initial convergence, i.e., evaluated DT sections are preceded by an STFE and an STNE section, while evaluated single-talk (ST) sections are preceded by a section of their respective condition. Each added section is $8$\,s to $12$\,s long and is removed before calculating metrics.

\begin{table}[t]
    \setlength{\tabcolsep}{.35em}
    \caption{Ablations on different knowledge distillation strategies for the {\tt CGGN16} student architecture based on the loss functions introduced in Section \ref{ssec:KD}, evaluated on the \textbf{dev set} $\mathcal{D}_{\mathrm{dev}}$ \textbf{double-talk condition}. Best results are {bold}, second best {underlined}. Two-step training is marked by the $\rightarrow$ symbol.}
    \centering
    \newcolumntype{R}{>{\raggedleft\arraybackslash}X}
\newcolumntype{C}{>{\center\arraybackslash}X}
\newcommand{\bftab}{\fontseries{b}\selectfont}

\begin{tabular}{l ccccc}
	\toprule
 \multirow{2}{*}{\textbf{Loss}} & \multicolumn{5}{c}{\textbf{Double-Talk (DT)}} \\

\cmidrule(lr){2-6}

& {PESQ} & {PESQ$_\mathrm{BB}$} & {ERLE$_\mathrm{BB}$} & {DT\,O} & {DT\,E} \\
\cmidrule(lr){1-6}

Unprocessed & 1.68 & 4.64 & -- & 3.88 & 1.96 \\

\cmidrule[0.75pt](lr){1-6}

${J}^\mathrm{GTt}$ \eqref{eq:GTt} & 2.15 & \underline{3.80} & \phantom{x}9.21 & 3.35 & 3.78 \\
${J}^\mathrm{GTKDt}$ \eqref{eq:GTKDt} & 2.22 & 3.62 & 11.91 & 3.40 & \underline{4.02}\\
${J}^\mathrm{GTKDf}$ \eqref{eq:GTKDf} & \textbf{2.27} & 3.71 & 10.90 & 3.32 & \underline{4.02}\\
${J}^\mathrm{KDf}$ \eqref{eq:KDf} & \underline{2.26} & 3.69 & \underline{11.97} & \underline{3.44} & 3.96\\
${J}^\mathrm{KDf}$ \eqref{eq:KDf} $\rightarrow {J}^\mathrm{GTt}$ \eqref{eq:GTt} & \textbf{2.27} & 3.67 & \textbf{12.78} & \textbf{3.49} & \textbf{4.04}\\
 
\bottomrule
\end{tabular}
    \label{tab:results_abla_dev}
\end{table}

\begin{table*}[t]
    \setlength{\tabcolsep}{.28em}
    \caption{Baseline and reference methods, along with the proposed {\tt CGGN16} in different sizes on the \textbf{test set} $\mathcal{D}_{\mathrm{test}}$. *-marked metrics are subjective MOS on noiseless data. Best results among -M/-S models are {bold}, second best {underlined}. Higher is better for all metrics apart LSD. Two-step KD training is marked by the $\rightarrow$ symbol.}
    \centering
    \newcolumntype{R}{>{\raggedleft\arraybackslash}X}
\newcolumntype{C}{>{\center\arraybackslash}X}
\newcommand{\bftab}{\fontseries{b}\selectfont}

\begin{tabular}{l r cc cccccccccc cc c}
	\toprule
 \multirow{2}{*}{\textbf{Method}} & \multirow{2}{*}{\hspace{-2mm}\textbf{Loss}} & & & \multicolumn{10}{c}{\textbf{Double-Talk (DT)}} & \multicolumn{2}{c}{\textbf{STFE}} & {\textbf{STNE}} \\

\cmidrule(lr){3-4}
\cmidrule(lr){5-14}
\cmidrule(lr){15-16}
\cmidrule(lr){17-17}

& & {{\#par.}} & {{\#FLOPS}} & {PESQ} & {PESQ$_\mathrm{BB}$} & {ERLE$_\mathrm{BB}$} & {DT\,O} & {DT\,E} & {DT\,O*} & {DT\,E*} & 
LPS & ESTOI & LSD$\downarrow$ &
{ERLE$_\mathrm{BB}$} & {ST\,E} & {PESQ} \\
\cmidrule(lr){1-17}

Unprocessed & & -- & -- & 1.68 & 4.64 & -- & 3.88 & 1.96 & -- & -- & 0.45 & 0.48 & 12.56 & -- & 1.92 & 2.62 \\

\cmidrule[0.75pt](lr){1-17}

{\tt FDKF}~\cite{Enzner2006, Franzen2018a} & & -- & \phantom{x}0.70 M & 1.85 & {4.35} & \phantom{x}3.04 & 3.67 & 2.49 & 2.75 & 3.09 & 0.61 & 0.53 & 11.08 & \phantom{x}3.53 & 2.52 & 2.62 \\
{\tt DLAC-Kalman}~\cite{Haubner2024} & \hspace{-10mm}\eqref{eq:GTt} & 50 k & \phantom{x}6.37 G & {2.09} & {3.57} & \phantom{x}9.92 & 3.70 & 3.86 & 3.04 & 3.43 & 0.73 & 0.58 & \phantom{x}9.69 & 12.04 & 3.96 & 2.62 \\ 
{\tt CRUSE-AEC}~\cite{Braun2022} & \hspace{-10mm}\eqref{eq:GTt} & 1.9 M & \phantom{x}0.82 G & 2.05 & 3.42 & 11.70 & 3.77 & 4.07 & 3.17 & 3.66 & 0.75 & 0.62 & \phantom{x}9.69 & 16.13 & 3.97 & 2.65 \\

\cmidrule[0.75pt](lr){1-17}

{\tt CGGN16-T} ($F\!=\!64, g\!=\!16$) & \hspace{-10mm}\eqref{eq:GTt} & {2.4} M & 12.47 G & {2.13} & {3.57} & {11.29} &{3.81} & {4.29} & 3.31 & 3.76 & 0.77 & 0.63 & \phantom{x}{9.60} & {16.89} & {4.35} & 2.68 \\

\cmidrule[0.01pt](lr){1-17}

{\tt CGGN16-M} ($F\!=\!24, g\!=\!6$) & \hspace{-10mm}\eqref{eq:GTt} & \underline{0.7 M} & \phantom{x}\underline{1.87 G} & \underline{2.04} & 3.51 & 10.65 & \textbf{3.79} & \textbf{4.18} & 2.90 & 3.47 & \underline{0.73} & \textbf{0.61} & \phantom{x}\underline{9.94} & \underline{15.46} & \textbf{4.08} & 2.64\\

{\tt CGGN16-S} ($F\!=\!8, g\!=\!2$) 
& \hspace{-10mm}\eqref{eq:GTt} & \textbf{0.2 M} & \textbf{\phantom{x}0.25 G} & 1.95 & 3.45 & \underline{10.68} & 3.65 & 3.93 & 3.03 & {3.57} & 0.70 & \underline{0.58} & 10.19 & 14.43 & 3.70 & 2.66\\
\ + w/ single-stage KD & \hspace{-10mm}\eqref{eq:KDf} & \textbf{0.2 M} & \textbf{\phantom{x}0.25 G} & \textbf{2.07} & \textbf{3.65} & \phantom{x}9.72 & \underline{3.73} & 4.07 & \textbf{3.17} & \textbf{3.61} & \textbf{0.74} & \textbf{0.61} & \phantom{x}\textbf{9.75} & 13.99 & 4.03 & \underline{2.67} \\
\ + w/ two-stage KD & \hspace{-10mm}\eqref{eq:KDf}$\rightarrow $\eqref{eq:GTt} & \textbf{0.2 M} & \textbf{\phantom{x}0.25 G} & \textbf{2.07} & \underline{3.56} & \textbf{11.19} & \textbf{3.79} & \underline{4.14} & \underline{3.11} & \underline{3.59} & \textbf{0.74} & \textbf{0.61} & \phantom{x}{9.95} & \textbf{15.59} & \underline{4.07} & \textbf{2.72} \\

\bottomrule
\end{tabular}
    \label{tab:results_final_test}
\end{table*}

\subsection{Ablations on Knowledge Distillation Strategy}
\label{ssec:dev}

Table \ref{tab:results_abla_dev} shows the results on $\mathcal{D}_{\mathrm{dev}}$ of employing our various loss functions introduced in Section \ref{ssec:KD} for the small {\tt CGGN16} student model using only $F=8$ base kernels and $g=2$ GRU groups. The first loss ${J}^\mathrm{GTt}$ \eqref{eq:GTt} represents the usual training on ground-truth labels, while all other experiments incorporate knowledge distillation. The implemented two-step training scheme is represented by ${J}^\mathrm{KDf} \rightarrow {J}^\mathrm{GTt}$. 

We can see that almost all KD losses improve performance over the ground-truth loss ${J}^\mathrm{GTt}$~\eqref{eq:GTt}. Mixing ground truth and time-domain KD in loss ${J}^\mathrm{GTKDt}$~\eqref{eq:GTKDt} yields a moderate improvement, while ${J}^\mathrm{GTKDf}$~\eqref{eq:GTKDf} with its frequency-domain KD loss component turned out more potent. The best performance among the single-step KD approaches (3 times in top rank) is achieved using  ${J}^\mathrm{KDf}$, completely omitting any ground-truth labels in training. Fine-tuning this model on ground-truth labels in a second step (\mbox{${J}^\mathrm{KDf} \rightarrow {J}^\mathrm{GTt}$}) improves the performance even further, achieving 4 times overall 1$^\text{st}$ rank. Interestingly, attempts to employ a more complex feature matching loss based on \cite{Nathoo2024} which also aims to align the latent features after the encoder (${\bf z}^\mathrm{ENC}$) and after the recurrent bottleneck (${\bf z}^\mathrm{rBN}$, cf.\ Fig.\ \ref{fig:CGGN}) between teacher and student 
failed to improve over the baseline w/o KD. The {\tt CGGN16} is seemingly not deep enough to take benefit from intermediate feature alignment. Also, the knowledge transfer from the much larger teacher might be more effective when the student model has more flexibility in applying this knowledge to its limited resources. We also observed that models employing KD often converged faster and needed less training time to reach their final performance.

\subsection{Test Set Results and Discussion}
\label{ssec:test}

Table \ref{tab:results_final_test} shows the results of our final proposed models with single-step \eqref{eq:KDf} and two-step \eqref{eq:KDf}$\rightarrow$\eqref{eq:GTt} knowledge distillation in comparison to larger {\tt CGGN16} variants as reference (\textbf{T}eacher, \textbf{M}edium, \textbf{S}tudent) as well as baseline methods from literature, including the classical {\tt FDKF}~\cite{Enzner2006, Franzen2018a}, Microsoft's CRN model {\tt CRUSE-AEC}~\cite{Braun2022}, and the classical/DNN hybrid {\tt DLAC-Kalman} model~\cite{Haubner2024}. The STNE condition serves as a sanity check: Models are not expected to improve PESQ scores a lot (as noise suppression was not part of the trained task), but should not degrade compared to unprocessed signals.

The results on \textit{instrumental} metrics for the models trained with KD 
follow the previous findings on the dev set, aside from the ${J}^\mathrm{KDf}$ \eqref{eq:KDf} training step achieving a lower ERLE$_{\rm BB}$ score in favor of better speech preservation. The two-step KD approach significantly excels {\tt CGGN16-S},
as both NE-related (PESQ$_{\rm BB}$, LPS, ESTOI, DT\,O) as well as echo-related metrics (ERLE$_{\rm BB}$, DT\,E, ST\,E) improve in double-talk and single-talk conditions.
\textit{Interestingly, we achieve about the same overall performance as the much larger \mbox{{\tt CGGN16-M}} model trained on ground-truth labels, despite using 70\% less parameters and only 14\% (i.e., a sixth) of the computational complexity.}

\textit{Subjective} metrics from our crowd-sourced P.808/P.831 listening test largely confirm these observations, but favor the smaller {\tt CGGN16-S} over the {\tt CGGN16-M}, especially after the first KD step \eqref{eq:KDf}. In DT\,O*, we can see a noticeable improvement over the model w/o KD. Overall, the models often show residual echo with noise characteristics, which typical postfilters can handle very well. Models with KD reduce the amount of distracting residual echo while also achieving higher NE speech quality in cases of an overpowering echo component in the microphone signal.
When comparing our overall subjectively best {\tt CGGN16} with KD \mbox{${J}^\mathrm{KDf}$} \eqref{eq:KDf} to baseline {\tt FDKF}~\cite{Enzner2006, Franzen2018a}, the use of a DNN helps also in PESQ and significantly in echo performance. Compared to {\tt DLAC-Kalman}~\cite{Haubner2024}, both of our KD approaches have less than 4\% of computational requirements, but still deliver better 
DT\,O* and DT\,E*
MOS scores. Compared to Microsoft's {\tt CRUSE-AEC}~\cite{Braun2022}, we achieve overall comparable performance at significantly less parameters and complexity.

\begin{figure}
    \hspace{-2.2em}%
        	\usetikzlibrary{dsp}
	\usetikzlibrary{chains}
	\usetikzlibrary{shapes.geometric}
	\usetikzlibrary{calc}
	\usetikzlibrary{arrows.meta}

\pgfkeys{
    /prepare label/.style={
        /print label/\detokenize{#1}/.code={\ttfamily\bf\scriptsize\detokenize{#1}}
    },
    /prepare label/.list={{w/o KD}, {w/ KD},CGGN-S,CGGN-M, CGGN-T, CRUSE, DLAC-Kalman, FDKF, -AEC, -Kalman}
}

\begin{tikzpicture}

\begin{axis}[
width=.60*\columnwidth,
height=.55*\columnwidth,
legend cell align={left},
legend style={fill opacity=1.0, draw opacity=1, text opacity=1, draw=white!80!black},
log basis x={10},
tick align=outside,
tick pos=left,
xlabel={\#FLOPS},
x label style={yshift=3mm},
xmin=110000000, xmax=20000000000,
xmode=log,
tick style={color=black},
y label style={anchor=north, xshift=19mm, yshift=-2.5mm},
ylabel={\rotatebox[origin=c]{270}{{DT\,O* (MOS speech quality)}}},
ymajorgrids,
yminorgrids,
minor y tick num=4,
ymin=2.70, ymax=3.349,
extra y ticks={4.6},
extra y tick labels={
    {}
},
extra y tick style={%
    major tick length=3pt,
},
extra x ticks={150000000},
extra x tick style={%
    every tick/.append style={white, line width=2.7pt},
    yshift=1pt,
    grid=none,
},
extra x tick labels={
    {}
},
]

\addplot[
        mark=o,
        line width=0.75pt,
        only marks,
        point meta=explicit symbolic,
        green,
        nodes near coords={
            \pgfkeys{/print label/\pgfplotspointmeta/.try}
        },
        every node near coord/.append style={font=\small, 
        anchor=south,yshift=-2.5pt, xshift=15pt}
    ]
        table[header=false,meta index=0,x index=1,y index=2]{
			{w/ KD} 250000000 3.11
        };
\addplot[
        line width=0.75pt,
        only marks,
        point meta=explicit symbolic,
        teal,
        nodes near coords={
            \pgfkeys{/print label/\pgfplotspointmeta/.try}
        },
        every node near coord/.append style={font=\small, 
        anchor=south,yshift=2.5pt, xshift=3.5pt}
    ]
        table[header=false,meta index=0,x index=1,y index=2]{
		      CGGN-S 250000000 3.2
        };
\addplot[
        mark=o,
        line width=0.75pt,
        only marks,
        point meta=explicit symbolic,
        teal,
        nodes near coords={
            \pgfkeys{/print label/\pgfplotspointmeta/.try}
        },
        every node near coord/.append style={font=\small, 
        anchor=south,yshift=1.0pt, xshift=2.5pt}
    ]
        table[header=false,meta index=0,x index=1,y index=2]{
			{w/ KD} 250000000 3.17
        };
        
\draw (axis cs:550000000,3.095) node (t) {\color{green}\bf\scriptsize(3)\,$\rightarrow$\,(1)};
\draw (axis cs:520000000,3.215) node (t) {\color{teal}\bf\scriptsize(3)};

\draw (axis cs:175000000,3.075) node (t) [fill=white] {\bf\tiny\color{green}KD};

\draw [arrows = {-Stealth[green, length=6pt]}, green, line width = 1.4pt] (axis cs: 250000000, 3.05) -- (axis cs: 250000000, 3.17);

\addplot[
        mark=star,
        line width=0.75pt,
        only marks,
        point meta=explicit symbolic,
        green,
        nodes near coords={
            \pgfkeys{/print label/\pgfplotspointmeta/.try}
        },
        every node near coord/.append style={font=\small, fill=white,
        anchor=south,yshift=-13.5pt, xshift=2pt}
    ]
        table[header=false,meta index=0,x index=1,y index=2]{
			CGGN-S 250000000 3.03
        };
\addplot[
        mark=star,
        line width=0.75pt,
        only marks,
        point meta=explicit symbolic,
        cyan,
        nodes near coords={
            \pgfkeys{/print label/\pgfplotspointmeta/.try}
        },
        every node near coord/.append style={font=\small, fill=white,
        anchor=south, yshift=-11pt, xshift=0pt, inner sep=2}
    ]
        table[header=false,meta index=0,x index=1,y index=2]{
			CGGN-M 1870000000 2.90
        };
\addplot[
        mark=star,
        line width=0.75pt,
        only marks,
        point meta=explicit symbolic,
        blue,
        nodes near coords={
            \pgfkeys{/print label/\pgfplotspointmeta/.try}
        },
        every node near coord/.append style={font=\small, fill=white,
        anchor=south,yshift=-11pt, xshift=-9pt, inner sep=2}
    ]
        table[header=false,meta index=0,x index=1,y index=2]{
			CGGN-T 12470000000 3.31
        };
\addplot[
        mark=star,
        line width=0.75pt,
        only marks,
        point meta=explicit symbolic,
        purple,
        nodes near coords={
            \pgfkeys{/print label/\pgfplotspointmeta/.try}
        },
        every node near coord/.append style={font=\small, fill=white, anchor=south,yshift=-12.0pt,
        xshift=-1.8pt, inner sep=2}
    ]
        table[header=false,meta index=0,x index=1,y index=2]{
			DLAC-Kalman 6370000000 3.04
        };
\addplot[
        mark=star,
        line width=0.75pt,
        only marks,
        point meta=explicit symbolic,
        orange,
        nodes near coords={
            \pgfkeys{/print label/\pgfplotspointmeta/.try}
        },
        every node near coord/.append style={font=\small, fill=white,
        anchor=south,yshift=1pt, xshift=13pt, inner sep=2}
    ]
        table[header=false,meta index=0,x index=1,y index=2]{
			FDKF 120000000 2.75
        };
\addplot[
        line width=0.75pt,
        only marks,
        point meta=explicit symbolic,
        magenta,
        nodes near coords={
            \pgfkeys{/print label/\pgfplotspointmeta/.try}
        },
        every node near coord/.append style={font=\small, anchor=south,yshift=0.5pt, xshift=15pt, fill=white, inner sep=1pt}
    ]
        table[header=false,meta index=0,x index=1,y index=2]{
			-AEC 820000000 3.143
        };
\addplot[
        mark=star,
        line width=0.75pt,
        only marks,
        point meta=explicit symbolic,
        magenta,
        nodes near coords={
            \pgfkeys{/print label/\pgfplotspointmeta/.try}
        },
        every node near coord/.append style={font=\small, anchor=south,yshift=3pt, xshift=15pt, fill=white, inner sep=1pt}
    ]
        table[header=false,meta index=0,x index=1,y index=2]{
			CRUSE 820000000 3.17
        };
    
\end{axis}

\draw (0.2,-0.15) node (t) {\rotatebox[origin=c]{270}{{\!\!\!\!\!\! $\approx$}}};

\end{tikzpicture}
    \hspace{-1.4em}%
        	\usetikzlibrary{dsp}
	\usetikzlibrary{chains}
	\usetikzlibrary{shapes.geometric}
	\usetikzlibrary{calc}
	\usetikzlibrary{arrows.meta}

\pgfkeys{
    /prepare label/.style={
        /print label/\detokenize{#1}/.code={\ttfamily\bf\scriptsize\detokenize{#1}}
    },
    /prepare label/.list={{w/ KD},CGGN-S,CGGN-M, CGGN-T, CRUSE, DLAC, -Kalman, FDKF, -AEC, -Kalman}
}

\begin{tikzpicture}

\definecolor{color0}{rgb}{0.67843137254902,0.847058823529412,0.901960784313726}

\begin{axis}[
width=.60*\columnwidth,
height=.55*\columnwidth,
legend cell align={left},
legend style={fill opacity=1.0, draw opacity=1, text opacity=1, draw=white!80!black},
log basis x={10},
tick align=outside,
tick pos=left,
x label style={yshift=3.5mm},
xlabel={\#parameters},
xmin=70000, xmax=3500000,
xmode=log,
tick style={color=black},
y label style={anchor=north, xshift=19mm, yshift=-2.5mm},
ylabel={\rotatebox[origin=c]{270}{{LPS (phonetic fidelity)}}},
ymajorgrids,
yminorgrids,
minor y tick num=4,
ymin=0.60, ymax=0.799,
extra x ticks={83000},
extra x tick style={%
    every tick/.append style={white, line width=2.7pt},
    yshift=1pt,
    grid=none,
},
extra x tick labels={
    {}
},
]
\addplot[
        line width=0.75pt,
        only marks,
        point meta=explicit symbolic,
        green,
        nodes near coords={
            \pgfkeys{/print label/\pgfplotspointmeta/.try}
        },
        every node near coord/.append style={font=\small, anchor=south, yshift=5.0pt, 
        xshift=0pt, fill=white, inner sep=2}
    ]
        table[header=false,meta index=0,x index=1,y index=2]{
			CGGN-S 220000 0.751
        };
\addplot[
        mark=o,
        line width=0.75pt,
        only marks,
        point meta=explicit symbolic,
        green,
        nodes near coords={
            \pgfkeys{/print label/\pgfplotspointmeta/.try}
        },
        every node near coord/.append style={font=\small, anchor=south,yshift=2.25pt, xshift=-11.5pt, fill=white, inner sep=2}
    ]
        table[header=false,meta index=0,x index=1,y index=2]{
			{w/ KD} 200000 0.74
        };
\addplot[
        mark=o,
        line width=0.75pt,
        mark size=1.25pt,
        only marks,
        point meta=explicit symbolic,
        teal,
        nodes near coords={
            \pgfkeys{/print label/\pgfplotspointmeta/.try}
        },
        every node near coord/.append style={font=\small, 
        anchor=south,yshift=-5.5pt, xshift=15pt}
    ]
        table[header=false,meta index=0,x index=1,y index=2]{
			{w/ KD} 200000 0.74
        };
\draw (axis cs:340000,0.755) node (t) [fill=white] {\color{green}\bf\scriptsize(3)\,$\rightarrow$\,(1)};
\draw (axis cs:630000,0.741) node (t) {\color{teal}\bf\scriptsize(3)};

\draw (axis cs:260000,0.717) node (t) [fill=white] {\bf\tiny\color{green}KD};

\draw [arrows = {-Stealth[green, length=5.5pt]}, green, line width = 1.4pt] (axis cs: 200000, 0.7065) -- (axis cs: 200000, 0.735);

\addplot[
        mark=star,
        line width=0.75pt,
        only marks,
        point meta=explicit symbolic,
        green,
        nodes near coords={
            \pgfkeys{/print label/\pgfplotspointmeta/.try}
        },
        every node near coord/.append style={font=\small, anchor=south,yshift=-12pt, fill=white, inner sep=2}
    ]
        table[header=false,meta index=0,x index=1,y index=2]{
			{CGGN-S} 200000 0.70
        };
\addplot[
        line width=0.75pt,
        only marks,
        point meta=explicit symbolic,
        purple,
        nodes near coords={
            \pgfkeys{/print label/\pgfplotspointmeta/.try}
        },
        every node near coord/.append style={font=\small, anchor=south,yshift=-10pt, xshift=5pt, fill=white, inner sep=1pt}
    ]
        table[header=false,meta index=0,x index=1,y index=2]{
			-Kalman 77000 0.7175
        };
\addplot[
        mark=star,
        line width=0.75pt,
        only marks,
        point meta=explicit symbolic,
        purple,
        nodes near coords={
            \pgfkeys{/print label/\pgfplotspointmeta/.try}
        },
        every node near coord/.append style={font=\small, anchor=south,yshift=-10pt, xshift=5pt, fill=white, inner sep=1pt}
    ]
        table[header=false,meta index=0,x index=1,y index=2]{
			DLAC 77000 0.73
        };

\addplot[
        mark=star,
        line width=0.75pt,
        only marks,
        point meta=explicit symbolic,
        cyan,
        nodes near coords={
            \pgfkeys{/print label/\pgfplotspointmeta/.try}
        },
        every node near coord/.append style={font=\small, anchor=south,yshift=-12pt, fill=white, inner sep=2}
    ]
        table[header=false,meta index=0,x index=1,y index=2]{
			CGGN-M 700000 0.73
        };
\addplot[
        mark=star,
        line width=0.75pt,
        only marks,
        point meta=explicit symbolic,
        blue,
        nodes near coords={
            \pgfkeys{/print label/\pgfplotspointmeta/.try}
        },
        every node near coord/.append style={font=\small, anchor=south,yshift=3pt, xshift=-9.5pt, fill=white, inner sep = 1pt}
    ]
        table[header=false,meta index=0,x index=1,y index=2]{
			CGGN-T 2400000 0.77
        };
\addplot[
        mark=star,
        line width=0.75pt,
        only marks,
        point meta=explicit symbolic,
        orange,
        nodes near coords={
            \pgfkeys{/print label/\pgfplotspointmeta/.try}
        },
        every node near coord/.append style={font=\small, anchor=south,yshift=2pt, xshift=10pt, fill=white, inner sep=1pt}
    ]
        table[header=false,meta index=0,x index=1,y index=2]{
			FDKF 75000 0.61
        };
\addplot[
        mark=star,
        line width=0.75pt,
        only marks,
        point meta=explicit symbolic,
        magenta,
        nodes near coords={
            \pgfkeys{/print label/\pgfplotspointmeta/.try}
        },
        every node near coord/.append style={font=\small, anchor=south,yshift=-10.5pt, xshift=0pt, fill=white, inner sep=1pt}
    ]
        table[header=false,meta index=0,x index=1,y index=2]{
			-AEC 1900000 0.737
        };
\addplot[
        mark=star,
        line width=0.75pt,
        only marks,
        point meta=explicit symbolic,
        magenta,
        nodes near coords={
            \pgfkeys{/print label/\pgfplotspointmeta/.try}
        },
        every node near coord/.append style={font=\small, anchor=south,yshift=-10.5pt, xshift=0pt, fill=white, inner sep=1pt}
    ]
        table[header=false,meta index=0,x index=1,y index=2]{
			CRUSE 1900000 0.75
        };

\end{axis}

\draw (0.15,-0.15) node (t) {\rotatebox[origin=c]{270}{\!\!\!\!\!\!{ $\approx$}}};

\end{tikzpicture}

    \caption{Model performance w.r.t.\ DT\,O* over \#FLOPS (left) and LPS over \#parameters (right), evaluated on the DT portion of \textbf{test set} $\mathcal{D}_\mathrm{test}$. The loss for KD is our proposed one-step ${J}^\mathrm{KDf}$ or two-step approach ${J}^\mathrm{KDf} \rightarrow {J}^\mathrm{GTt}$.}
    \label{fig:complexity_plot}
\end{figure}
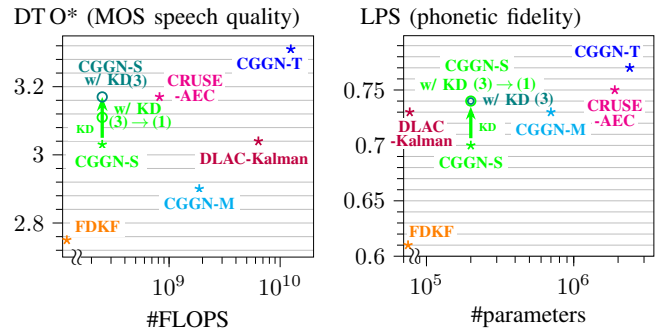

The improvements to the model performance gained from our KD models are also clearly visible in Fig.\ \ref{fig:complexity_plot}, where the exemplary DT\,O* and LPS scores are plotted against computational complexity and parameter count of the models, respectively. On the one hand, we can see the overall trend of (slightly) degraded performance of smaller and less complex models. On the other hand, we observe that a large portion of this performance decrease can be regained for the {\tt CGGN16-S} by KD, with the distilled students even clearly outperforming the much larger and more complex {\tt CGGN-M} in both metrics. Again, the performance of our proposed KD models is largely similar to the bigger and more complex {\tt CRUSE-AEC}.

\section{Conclusions}
\label{sec:results}

In this paper, we have shown how the application of knowledge distillation (KD) can significantly mitigate the performance loss from model downscaling for the task of acoustic echo control (AEC) and consequently enables both efficient and high-performance AEC models. The adopted teacher-student learning approach provides models which achieve improved near-end speech preservation and yield a clearly better overall performance over a much larger model trained on ground-truth labels, while using only 14\% of the FLOPS.

\bibliographystyle{IEEEtran}
\bibliography{refs, ifn_spaml_bibliography}

\end{document}